\documentclass{npw}
\usepackage{graphicx}

\begin{document}

\title{Microscopic description of rotation: \\
From ground states to the extremes of ultra-high spin}
%
%
\author{
  A.\ V.\ Afanasjev\email{afansjev@erc.msstate.edu} \\
  \it Department of Physics and Astronomy, Mississippi State
University, MS 39762}
%
\date{}
\maketitle

\begin{abstract}
 Recent progress in the microscopic description of rotational
properties within covariant density functional theory (CDFT) 
is presented. It is shown that it provides an accurate description
of rotational bands both in the paired regime at low spin and in
the unpaired regime at ultra-high spins. The predictive power of 
CDFT is verified by comparing the CDFT predictions for band 
crossing features in the $A\geq 242$ actinides with new experimental 
data. In addition, possible role of the Coulomb antipairing effect for 
proton pairing is discussed.
\end{abstract}

\section{Introduction}

Low-energy theoretical nuclear physics aims at the description 
of different aspects of the nuclear many-body problem for a wide 
variety of nuclei ranging from the proton to the neutron drip line 
\cite{E.12,AARR.13} and beyond. There is also a strong correlation 
between the advent of new experimental facilities and theoretical 
developments supporting new physics studied by these facilities 
at the limits of charge, isospin, spin, deformation etc.  The 
investigation of rotating nuclei is an important avenue for these 
studies  at the limits. For example, rotational properties serve 
as an important tool for configuration assignments in odd-mass light 
superheavy nuclei \cite{HG.08,AO.13}. The stability of nuclei 
against  fission at high spin \cite{ER.00} and the role of 
proton-neutron pairing \cite{FS.99,A.12} can also be addressed by studying 
rotating  nuclei.

  So far, the majority of theoretical studies of rotating 
nuclei have been performed within phenomenological approaches based on 
the Nilsson or Woods-Saxon potentials. Alternative and more microscopic 
approaches are based on nonrelativistic \cite{BHR.03} and relativistic 
(covariant) \cite{VALR.05} density functional theories (DFT); the latter
is usually called covariant DFT (CDFT). Until recently, these approaches 
were only occasionally used for the description of rotational structures
in the pairing regime and no systematic assessment of their errors and 
the sources of these errors is available. They were mostly applied
to superdeformed (SD) rotational bands in different mass regions 
(see Refs.\ \cite{AO.13,VALR.05} and references therein), the spins 
and parities of which are not known in most cases.
Prior to Ref.\ \cite{AO.13}, the studies of the normal-deformed (ND) 
bands in even-even and especially odd-mass nuclei over the observable 
frequency ranges have been performed only in a few nuclei (see Ref.\ 
\cite{AO.13} for full list of studied nuclei).

  The systematic investigation \cite{AO.13} of normal-deformed 
rotational bands in even-even and odd-mass actinides and light 
superheavy nuclei within the CDFT framework, performed for the first 
time in the density functional theory framework, fills significant 
gaps mentioned above in our  knowledge of the performance of microscopic 
theories. A short overview of these results with  the verification of 
the predictions of Ref.\ \cite{AO.13} is presented in Sect.\ 
\ref{actinides}. The latter is done by comparing the CDFT predictions 
with new data on rotational bands in Pu, Cm and Cf even-even nuclei. 
The role of the Coulomb antipairing effect for proton pairing is discussed 
in Sect.\ \ref{Coulomb-anti}. The performance of CDFT in the description 
of triaxial superdeformed (TSD) rotational bands at ultra-high spin is 
analyzed in Sect.\ \ref{158Er-ch}. Finally, the conclusions are 
presented in Sect.\ \ref{Concl}.

\section{Upbendings in actinides: confronting predictions with new
experimental data}
\label{actinides}

\begin{figure*}[ht]
\centering
\includegraphics[width=18cm,angle=0]{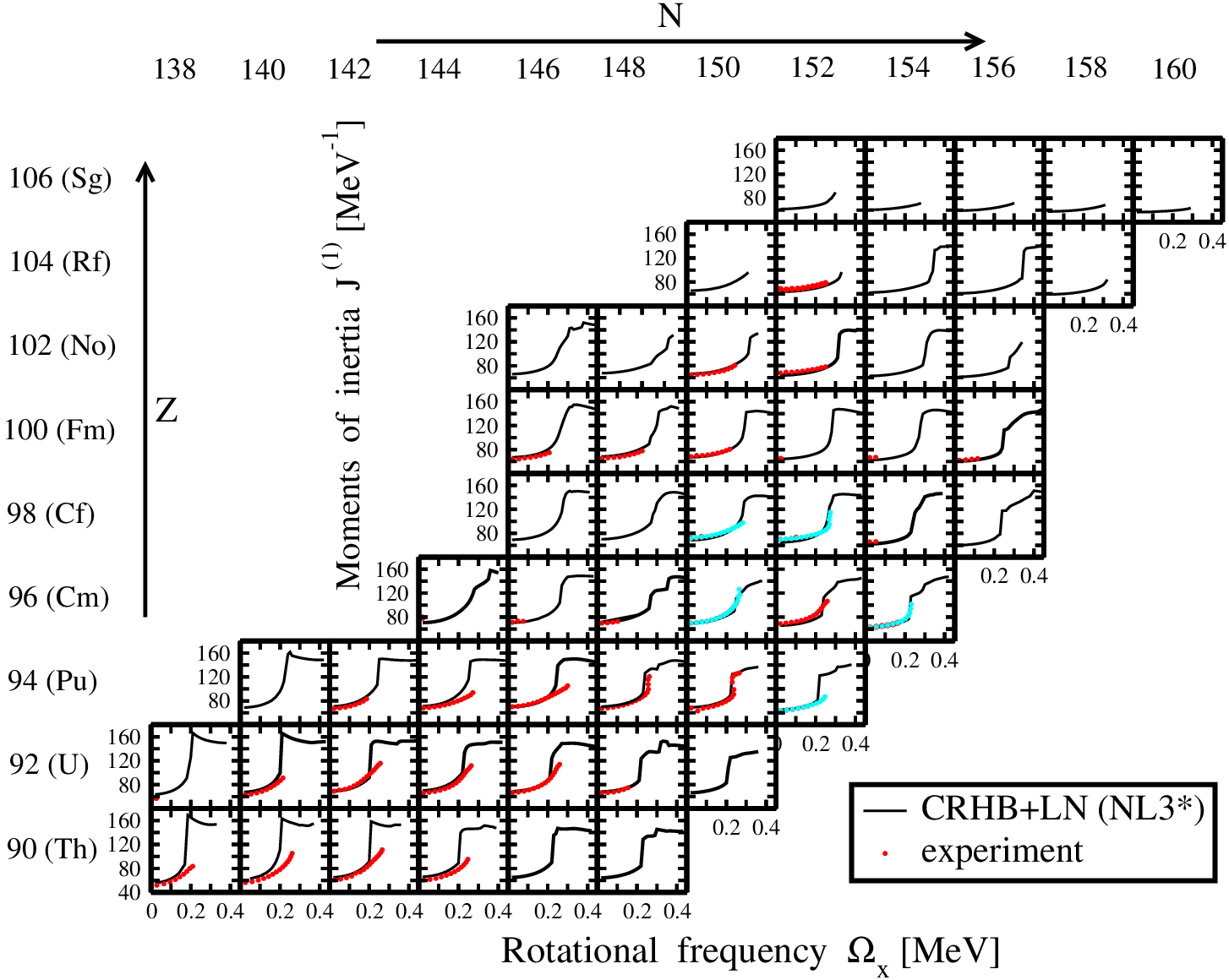}
\vspace{-0.8cm}
\caption{The experimental and calculated moments 
of inertia $J^{(1)}$ as a function of rotational frequency $\Omega_x$.
The calculations are performed with the NL3* parametrization 
\cite{NL3s} of CDFT. Calculated results and experimental data are 
shown by black lines and red dots, respectively. Cyan dots are
used for new data from Ref.\ \cite{Hota.thesis}.} 
\label{sys-J1-NL3s}
\end{figure*}

  Fig.\ \ref{sys-J1-NL3s} shows the results of the first ever (in 
any DFT framework) systematic investigation of rotational properties 
of even-even nuclei at normal deformation \cite{AO.13}. The calculations
are performed within the cranked relativistic Hartree-Bogoliubov
(CRHB) approach with approximate particle number projection by 
means of the Lipkin-Nogami method (further CRHB+LN) \cite{CRHB}. 
Cyan dots show new experimental data from Ref.\ \cite{Hota.thesis} 
which were not included in Ref.\ \cite{AO.13}. These data will be 
analyzed in detail in the current manuscript. One can see that the 
moments of inertia below band crossings are reproduced well. The 
upbendings observed in a number of rotational bands of the $A\geq 242$ 
nuclei are also reasonably well described in model calculations. 
However, the calculations also 
predict similar upbendings in lighter nuclei, but they have not
been seen in experiment. The analysis suggests that the stabilization 
of octupole deformation at high spin, not included in the present 
CRHB+LN calculations, could be responsible for this discrepancy 
between theory and experiment \cite{AO.13}. With few exceptions, the 
rotational properties of one-quasiparticle configurations, which yield 
important information on their underlying structure and, thus, provide 
an extra tool for configuration assignment, are also well described 
in the CRHB+LN calculations (see Ref.\ \cite{AO.13} for details).

 New experimental data on high-spin structures in even-even actinides
\cite{Hota.thesis}, not analyzed in Ref.\ \cite{AO.13}, include the 
ground state rotational bands in $^{246}$Pu, 
$^{246,250}$Cm and $^{248,250}$Cf (cyan dots in Fig.\ 
\ref{sys-J1-NL3s}). It should be kept in mind that one or several
of the highest spin transition(s) are tentative.
It is interesting to compare these data with the predictions of our 
CRHB+LN calculations \cite{AO.13} and the ones of the cranked shell 
model  with the pairing correlations treated by a particle-number 
conserving method (further CSM+PNC)
\cite{ZHZZZ.12}. The main difference between these two models is 
the treatment of the single-particle states. The NL1 \cite{NL1} 
and NL3* \cite{NL3s} parametrizations of the CDFT theory have been fitted 
only to bulk properties of around ten spherical nuclei [single-particle 
information such as the energies of 
single-particle states or spin-orbit splittings has not been used in the 
CDFT fits]. On the contrary, the parameters of the Nilsson potential were 
carefully adjusted to the experimental energies of deformed one-quasiparticle 
states of actinides in Ref.\ \cite{ZHZZZ.12}. Note that the predictions of 
the CSM+PNC model are available only for the Cm and Cf isotopes. Another 
major difference between the two models is the treatment of deformation. 
The deformations are chosen to be close to experimental values and do not 
change with rotational frequency in the CSM+PNC calculations, while 
equilibrium deformations are defined fully self-consistently at all 
calculated rotational frequencies in the CRHB+LN calculations. This 
procedure reveals considerable changes in quadrupole 
($\beta_2$) and triaxial ($\gamma$) deformations at the band crossing in the 
CRHB+LN calculations (see Fig.\ 7 in Ref.\ \cite{A250}) which are completely 
ignored in the  CSM+PNC calculations.

\begin{figure}[ht]
\includegraphics[width=8.7cm]{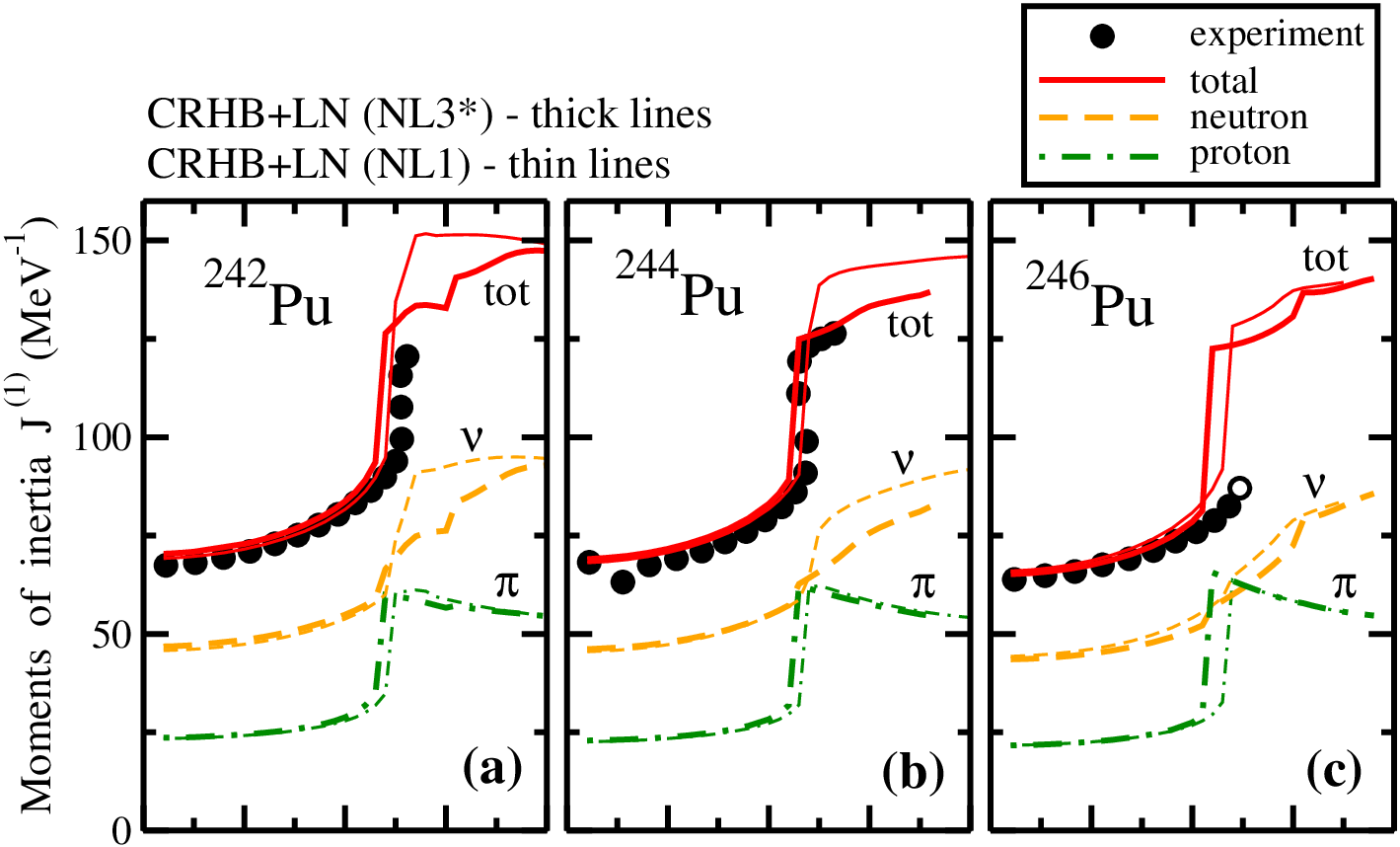}
\includegraphics[width=8.8cm]{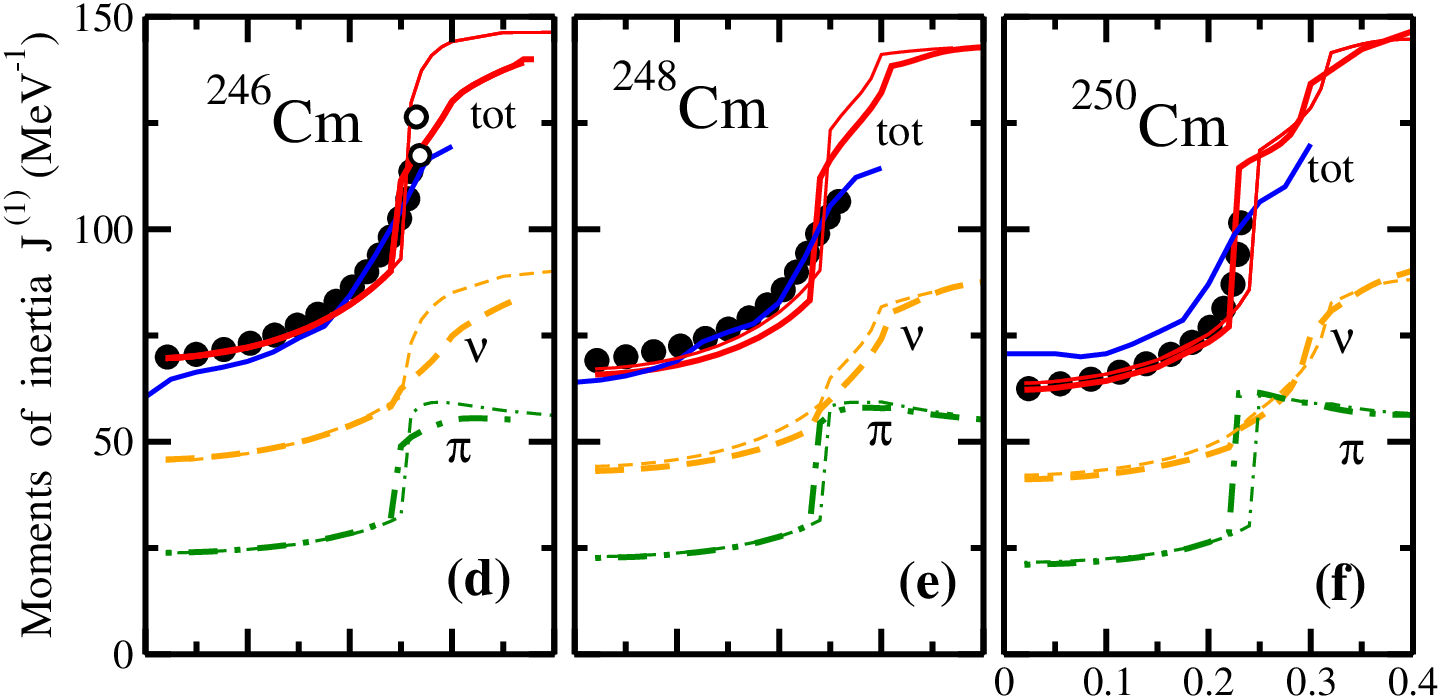}
\includegraphics[width=6.8cm]{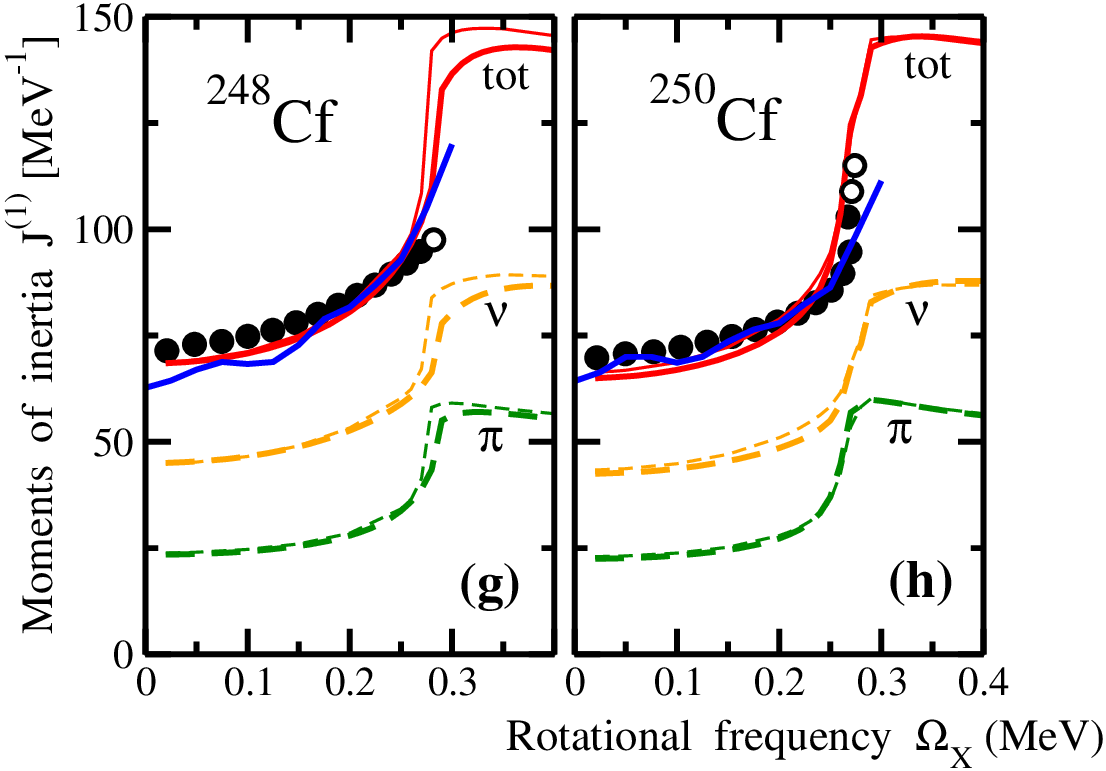}
\vspace{-0.4cm}
\caption{
The experimental and calculated kinematic moments of inertia $J^{(1)}$ 
of ground state rotational bands in indicated nuclei as a function of 
rotational frequency $\Omega_x$. 
Proton and neutron
contributions to the kinematic moment of inerta are presented. Open 
circles are used for tentative experimental points.} 
\label{back_Pu}
\end{figure}

Fig.\ \ref{back_Pu} compares CSM+PNC and CRHB+LN results with 
new experimental data of Ref.\ \cite{Hota.thesis}. The $^{242,244}$Pu 
data have already been analyzed in Sect.\ 4 of Ref.\ \cite{AO.13}, and 
are shown in panels (a) and (b) for completeness. The ground state rotational 
band in $^{246}$Pu has been extended to higher spins in Ref.\ 
\cite{Hota.thesis} as compared with earlier data. Its kinematic moment 
of inertia shows a rapid increase at the highest observed frequencies 
similar to the one seen before upbendings in $^{242,244}$Pu. However, the 
$^{246}$Pu data does not reveal an upbend yet. The CRHB+LN calculations 
predict an upbend in all three even-even Pu isotopes. The upbend in 
$^{242,244}$Pu is predicted $0.01-0.02$ MeV earlier in the CRHB+LN(NL3*) 
calculations as compared with experiment. A similar situation
is expected in $^{246}$Pu. Considering this and the fact that the last 
observed point in $^{246}$Pu is tentative, one can conclude that 
there is no significant discrepancies with experimental data. Even 
better agreement with this new data is seen in the case of the 
CRHB+LN(NL1) calculations.

 A smooth upbending takes place in $^{248}$Cm (Fig.\ \ref{back_Pu}e). 
It is rather well reproduced in the CSM+PNC calculations. The CRHB+LN 
calculations of Ref.\ \cite{AO.13} suggest that this upbending is 
predominantly due to the proton $i_{13/2}$ alignment. However, the 
interaction between the $g$ and $S$ bands in the band crossing region 
is too weak in the proton subsystem, which, in contradiction to 
experiment, leads to a sharp upbending in CRHB+LN calculations. 
A sharp upbending is also seen in the $J^{(1)}$ of $^{250}$Cm 
(Fig.\ \ref{back_Pu}f). It is well reproduced in the CRHB+LN calculations. 
In contrast, the CSM+PNC calculations show a gradual alignment in the 
band crossing region which contradicts the experiment. The upbending is 
also present in $^{246}$Cm (Fig.\ \ref{back_Pu}f). Both models account 
for this upbending, but differ in the details of its description. In the 
CRHB+LN  calculations, this upbend is somewhat sharper than in the 
experiment due to the sharp alignment of the $i_{13/2}$ proton pair. 
The CSM+PNC calculations seem to reproduce the gradual  character of 
this alignment better. However, they underestimate the alignment gain 
at the band crossing. The step in rotational frequency at which the 
CSM+NPC calculations are performed  is not specified in Ref.\ 
\cite{ZHZZZ.12}. If a large step is used, this may be a reason why 
CSM+NPC calculations look smoother in band crossing region than the 
CRHB+LN ones. The CRHB+LN calculations 
are performed in a step of 0.01 MeV in the band crossing region and, as a 
result, they reveal more details.

 The kinematic moment of inertia of the ground state band in $^{248}$Cf 
(see Fig.\ \ref{back_Pu}g) does not reveal an upbend  which is predicted 
both in the CRHB+LN and CSM+PNC calculations. However, considering that 
the experimental point at the  highest frequency  is tentative and the 
difference between calculations and experiment for the point before this 
one is not too large, a definite conclusion on whether the calculations 
fail to reproduce the experimental data is not possible. This is especially 
true considering that the CRHB+LN calculations tend to predict sharp 
upbends at lower (by 0.01-0.02 MeV) frequency as compared with experiment 
in this mass region (Fig.\ \ref{back_Pu}). The sharp upbend in $^{250}$Cf 
is well reproduced in the CRHB+LN calculations (see Fig.\ \ref{back_Pu}h). 
On the contrary, the CSM+PNC calculations predict a more gradual increase 
of $J^{(1)}$ with frequency which deviates more from experiment relatively 
to the CRHB+LN calculations.

  When comparing CRHB+LN and CSM+PNC calculations, one has to keep in 
mind that the former provides a much more consistent description of 
rotational motion in the paired regime. The pairing strength has been 
fitted to experiment in both approaches \cite{AO.13,ZHZZZ.12}. However, 
the pairing strength is different in even-even and odd-mass nuclei in 
the CSM+PNC approach \cite{ZHZZZ.12}; this is a well known deficiency of 
the cranked shell model (see, for example, Ref.\ \cite{CSM-exp}). In
contrast, the same pairing strength is used in even-even and odd-mass 
nuclei in the CRHB+LN approach and, according to Ref.\ \cite{AO.13}, it 
leads to a consistent and accurate description of odd-even mass staggerings 
(the $\Delta^{(3)}$ indicators) and the moments of inertia in even-even 
and odd-mass actinides. To our knowledge, this is achieved  in a 
systematic calculations for the  first time in the mean field/DFT based 
models. 

  A special effort has been made in the CSM+PNC approach to accurately
reproduce the energies of deformed one-quasiparticle states. Considering
that the strength of the interaction between the $g$ and $S$ bands and the 
crossing frequency depends sensitively on the relative position of aligning 
high-$j$ orbital with respect to the quasiparticle vacuum \cite{F.priv}, one
may think that this  will improve the description of band crossing properties. 
However, the comparison of the data with the results of the calculations 
shows that, on average, the CSM+PNC and CRHB+LN approaches describe observed 
upbendings with the same level of accuracy (Fig.\ \ref{back_Pu}).

  
  In contrast, single-particle information is not used at
all in the fits of the CDFT parametrizations \cite{NL3s,NL1}.
The accuracy of the description of experimental energies of 
one-quasiparticle deformed states is lower in CDFT as compared 
with the Nilsson potential (Ref.\ \cite{AS.11}), which, in particular,
is a consequence of the stretching out of the energy scale due to the
low effective mass of the nucleon. Despite these facts, the average 
accuracy of the description of rotational properties in the band 
crossing region is similar in both models. In addition, an accurate 
description of the rotational properties in odd-mass nuclei in CRHB+LN 
\cite{AO.13} is achieved in a more consistent way than in the CSM-NPC 
model (see discussion above). Whether the alignment in the band crossing 
region proceeds in a gradual (gradual increase of $J^{(1)}$) or sharp 
(sharp upbend in $J^{(1)}$) way depends on whether the interaction 
strength between the $g$ and $S$  bands is strong or 
weak. It follows from the results of both calculations and experiment 
that this interaction strength shows variations with particle number which 
are not always reproduced in both model calculations. This is where the 
differences between the CDFT parametrizations (NL1 and NL3*) and two 
models (CRHB+LN and CSM+PNC) with respect to the description of 
single-particle states show up. For example, they are responsible
for the differences in the alignments of the $j_{15/2}$ neutrons 
in $^{242,244}$Pu and $^{246}$Cm (Figs.\ \ref{back_Pu}a, b and d)
which align gradually (sharply) in the CRHB+LN calculations with
NL3* (NL1) parametrization. The high density of the single-particle
states in the actinides is another possible factor which decreases 
the predictive power of the models in the band crossing region.

\section{Coulomb antipairing effect for proton pairing}
\label{Coulomb-anti}

  The investigation of pairing and rotational properties of
actinides in Ref.\ \cite{AO.13} shows that the strengths of 
pairing defined by means of the moments of inertia and 
three-point $\Delta^{(3)}$ indicators strongly correlate
in the CRHB(+LN) framework. This allows to address the role 
of the Coulomb antipairing effect \cite{AER.01} in the 
description of pairing in the proton  subsystem.

\begin{figure}[h]
\centering
\includegraphics[width=\columnwidth]{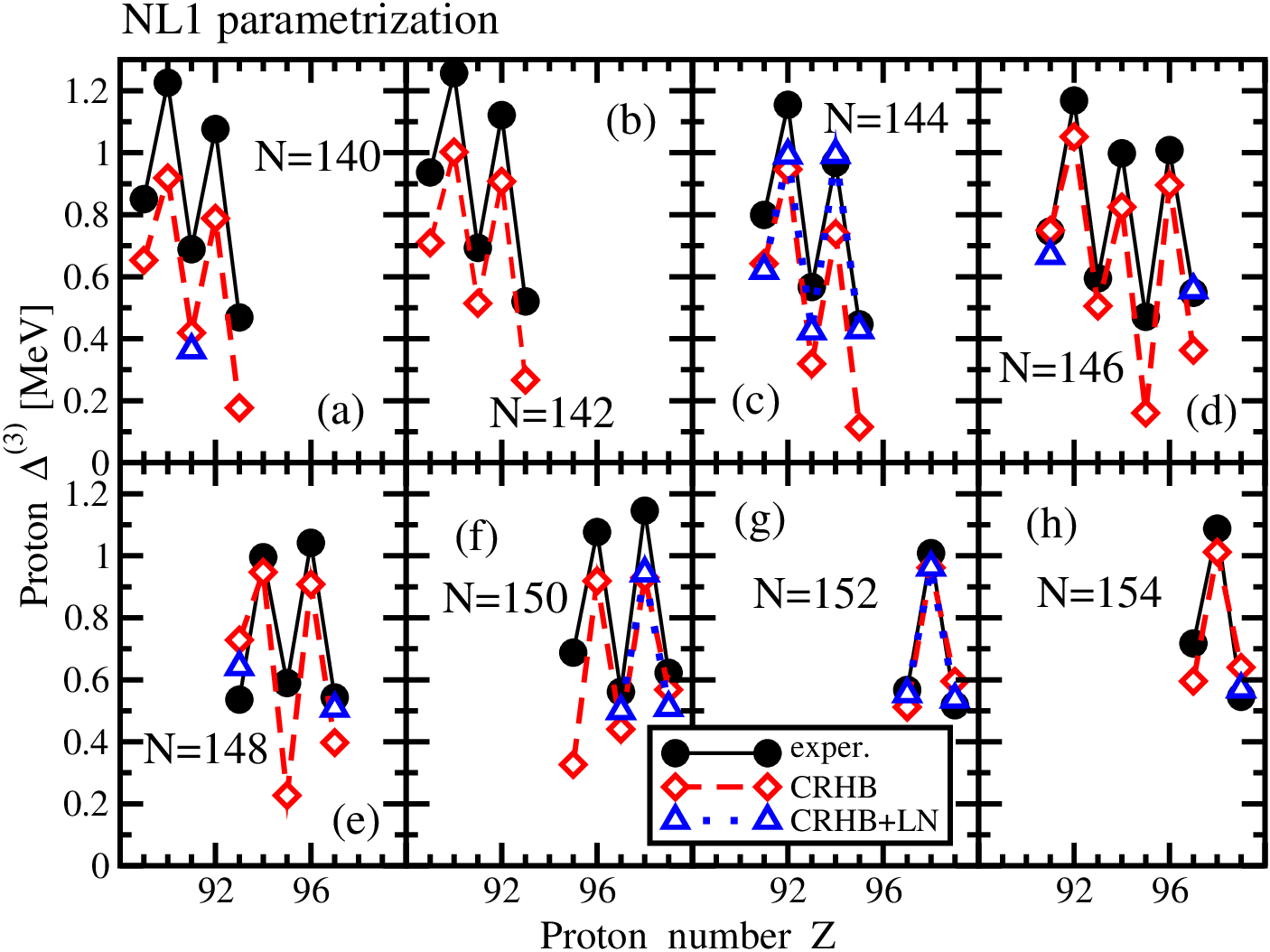}
\vspace{-0.8cm}
\caption{Experimental and calculated proton three-point 
indicators $\Delta^{(3)}_{\nu}(N)$ as a function of neutron number $N$. 
The results of the CRHB and CRHB+LN calculations with the NL1 
parametrization are shown. From Ref.\ \cite{AO.13}.}
\label{Delta3P-NL1}
\end{figure}

 The Coulomb force is not explicitly included into the pairing channel
of most DFT calculations because of its non-local nature. However, 
it is known that proton pairing gaps are reduced by 20-30\% if 
an exact Coulomb term is included into the calculations 
of the pairing field \cite{AER.01,LDBM.09,NY.11}; this term leads
to the so-called Coulomb antipairing effect \cite{AER.01}. 
Proton pairing energies and the moments of inertia of the proton 
subsystem are also strongly affected by it \cite{AER.01}. However, 
the Coulomb term is neglected in the calculations of the pairing 
field in the RHB framework \cite{VALR.05}.
Thus, it is important to understand to which extent our 
approach provides a correct description of proton pairing despite 
the fact that the Coulomb contribution to pairing is neglected. It 
turns out that the proton $\Delta^{(3)}_{\pi}$ indicators are 
correctly described in the CRHB+LN and CRHB calculations (see 
Fig.\ \ref{Delta3P-NL1} in the current manucript and Figs.\ 5, 7, 
and 8 in Ref.\ \cite{AO.13}).  The effect of the Coulomb 
interaction can be simulated by a renormalization scheme via a 
reduction factor of $\gamma_p=0.9$ \cite{NY.11} for the proton 
pairing channel. However, the CRHB and CRHB+LN calculations 
with this renormalization scheme lead to a frequent collapse of the 
proton pairing and to the proton $\Delta_{\pi}^{(3)}$ indicators 
which are too low as compared with experiment. Thus, the Brink-Booker 
part of the finite range Gogny D1S force, used in the pairing channel
of the CRHB(+LN) calculations, has to be treated as an 
effective pairing force without the Coulomb part, and as such it 
works rather well in the description of experimental proton pairing. 
As discussed in Ref.\ \cite{AER.01}, this is a consequence of fitting 
strategies of the Gogny force parametrizations which effectively 
neglect the Coulomb term in the pairing channel. Similar to our results, 
this reference shows that the inclusion of the Coulomb term in 
the calculations of pairing worsens the good agreement of theoretical 
results with experimental data, reinforcing the conjecture that the 
current fits of the Gogny force should be used without Coulomb terms
in the pairing channel.

  It is also necessary to recognize that the Coulomb antipairing 
effect has not been proven experimentally due to the difficulties 
of disentangling this phenomenon from other effects in the 
$N\neq Z$ systems. However, even-even $N=Z$ systems provide 
an excellent laboratory to test this effect, since the similarity 
of the proton and neutron single-particle spectra (apart from 
some constant shift in absolute energies by the Coulomb energy) leads to 
the fact that proton and neutron pairing energies are almost the 
same for the proton and neutron subsystems in calculations which do not 
contain a Coulomb term in the pairing channel (as is the case with the 
CRHB+LN calculations). As a consequence, the alignment (paired band 
crossing) of proton and neutron pairs in the ground state rotational 
bands of the $N=Z$ nuclei takes place at the same rotational frequency 
in such calculations, which in turn leads to only one bump in the 
dynamic moment of inertia $J^{(2)}$ curve. The proton pairing energies 
would be substantially lower than the neutron ones in the calculations 
with the Coulomb term in the pairing channel \cite{AER.01}. As a result, 
the alignment of the proton pair is expected at higher frequency as 
compared with the neutron one, leading to a double peaked shape for the 
dynamic moments of inertia.  However, as discussed in detail in Ref.\ 
\cite{Sr76} (in particular, see Fig.\ 3 in Ref.\ \cite{Sr76} and its 
discussion) the currently available experimental data on ground state 
rotational bands in even-even $N = Z$ nuclei shows only one peak in $J^{(2)}$ 
originating from a paired band crossing. This does not support the 
existence of the Coulomb antipairing effect. One may argue that
the presence of isoscalar neutron-proton pairing could lead to a 
situation in which these two peaks in $J^{(2)}$ merge into one
which is broader because of the mixing caused by this type of pairing. 
However, currenly there are no strong evidence for the existence 
of isoscalar $np$-pairing, see review in Ref.\ \cite{A.12}.

\section{The extremes of ultra-high spins}
\label{158Er-ch}
 
  While the rotational bands of the actinides are mostly near-prolate,
triaxial superdeformed (TSD) bands represent another class of 
rotational structures built on static triaxial shapes. Of particular
interest are the bands recently observed at ultra-high spins in the 
$A \sim 154-160$ mass region \cite{Er157-58-exp}. The TSD bands 1 
and 2 in $^{158}$Er were recently studied in Ref.\ \cite{Er158} within 
the cranked relativistic mean field (CRMF) \cite{CRMF} theory, which 
represents the  limit of the CRHB theory for the case of no pairing 
\cite{VALR.05}. Similar to the actinides (Sect.\ \ref{actinides}), the 
NL1 and NL3* parametrizations of the CDFT were employed in this study.

\begin{figure}[h]
\centering
\includegraphics[width=\columnwidth]{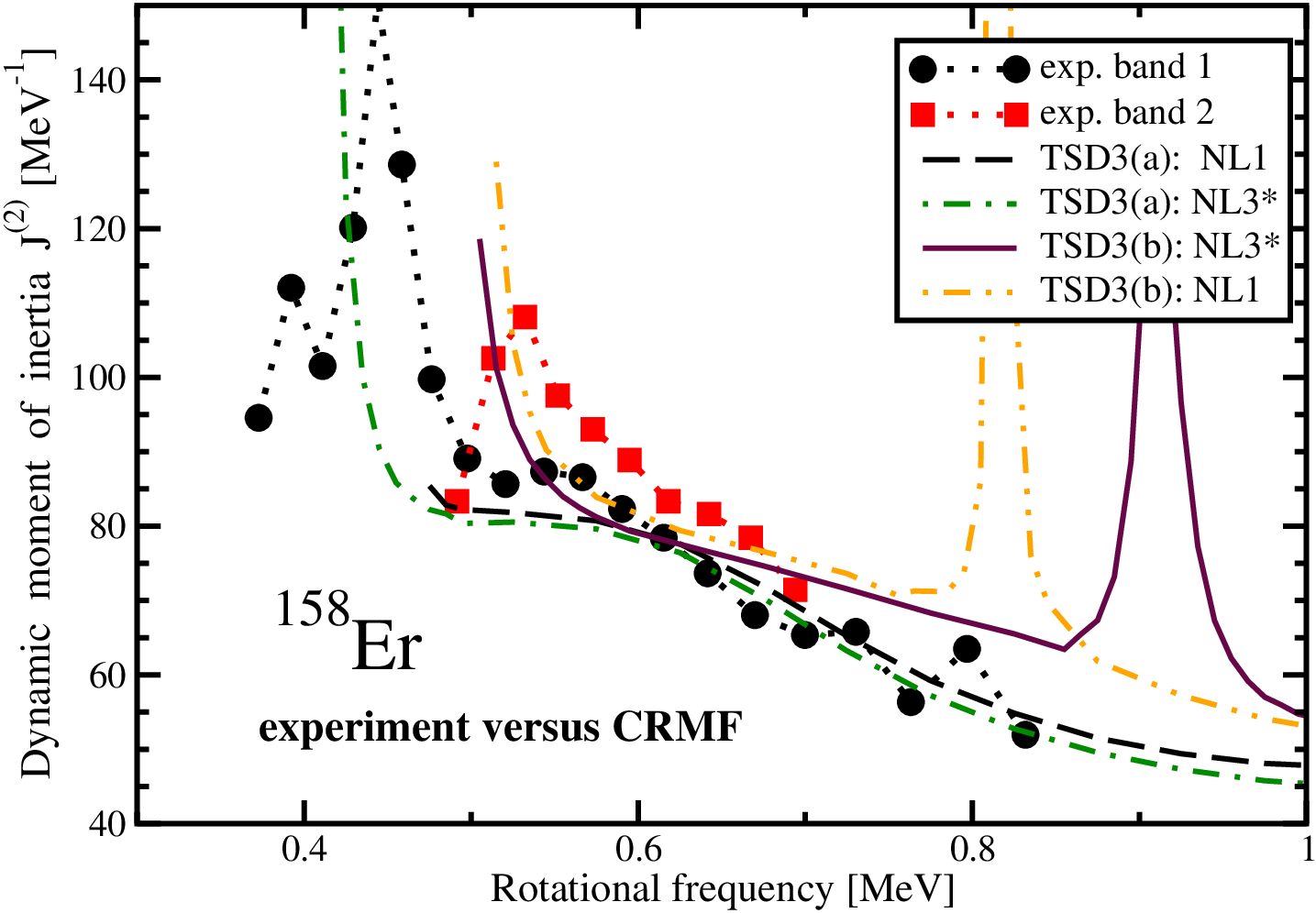}
\vspace{-0.8cm}
\caption{Experimental dynamic moments of inertia of observed 
TSD bands in $^{158}$Er (symbols) compared to calculated ones 
(lines). Based on results presented in Fig.\ 4 of Ref.\ 
\cite{Er158}.}
\label{J2-moments}
\end{figure}

  The degree of accuracy of the description of experimental data is 
illustrated in Fig.\ \ref{J2-moments}. The configuration TSD3(a), 
involving two protons in $N=6$ shell and one neutron in $N=7$ shell, 
i.e. $\pi 6^2 \nu 7^1$, is a possible candidate for the observed band 
1 in $^{158}$Er. The experimental dynamic moment of inertia $J^{(2)}$  
is rather well reproduced by assuming this configuration above the 
band crossing at low frequencies; the level of agreement with experiment 
is comparable to that obtained earlier for superdeformed  bands in the 
$A\sim 150$ region \cite{CRMF,CRMF-align}. Our CRMF-NL3* calculations 
suggest that the jump in dynamic moment of  inertia of band 1 at low 
frequencies can be associated with a band crossing with large interaction 
between the $1/2[770](r=+i)$ and $[N=5](r=+i)$ neutron routhians. 
The calculated transition quadrupole moment $Q_t$ of TSD3(a) changes 
from 10.5\,$e$b at $I=42$ to 9.0\,$e$b at $I=72$ and  $\gamma$ 
increases  slightly from  $12^{\circ}$ to $16^{\circ}$ in this spin range. 
Considering that the experimental value of $Q_t\sim 11$ $e$b 
\cite{Er158-qt} is subject to $\approx 15\%$  uncertainty due to nuclear 
and electronic stopping powers, these values are reasonably close to 
experiment. The comparison between experimental and calculated
energies shown in Ref.\ \cite{Er158} indicates that, to be consistent
with TSD3(a), band 1 has to be observed in the spin range
$I = 35-77$. If these theoretical spin assignments turned out to be 
correct, the experimental TSD band 1 in $^{158}$Er would be
the highest spin structure ever observed.

  If TSD3(a) is assigned to band 1, then the configuration TSD3(b) built 
upon TSD3(a) by exciting a neutron from $N=5(r = +i)$ into 
$5/2[642](r = -i)$ is a natural candidate for experimental band 2. The 
$J^{(2)}$ curve of this configuration is close to that of band 2 
(Fig.\ \ref{J2-moments}), and its transition quadrupole moment is only 
slightly larger (by $\sim 0.7$ $e$b) than that of TSD3(a) 
which is also close to experiment.

\section{Summary}
\label{Concl}

  The analysis of rotational spectra within the paired (CRHB+LN)
and unpaired (CRMF) cranking versions of the CDFT theory shows 
that the evolution of the moments of inertia as a function of 
rotational frequency, particle number and configuration [in odd-mass
nuclei] are well reproduced in model calculations both at low and at 
ultra-high spins. The inaccuracies in the description of the energies 
of one-quasiparticle states do not substantially affect the 
calculated moments of inertia outside the band crossing region.
Most of the observed upbendings are well reproduced in model 
calculations. However, in some cases, the details of the alignments 
(gradual alignment of $J^{(1)}$ or sharp upbend in $J^{(1)}$) in the 
band crossing region are not reproduced; this is true for both 
the CRHB+LN and CSM-NPC approaches. There may be two possible 
reasons for this observation. First, the interaction strength
between the $g$ and $S$ bands depends on fine details of the 
single-particle structure which are not reproduced in model
calculations with the required accuracy.
 Second, because of the limitations of the cranking
model in the band crossing region \cite{H.76}, methods beyond mean 
field may be required for a detailed description of band crossing
features in some cases.

  The current manuscript illustrates that the cranking 
model based on CDFT remains a powerful method for the study 
of rotating nuclei. In our experience, it works well when the mean 
field is well defined or the configuration interaction is weak; 
the latter takes place in high-spin structures with negligible 
pairing. The cranking model does not provide reliable results for 
nuclei with a very soft potential energy surfaces as examplified 
in Ref.\ \cite{Mo}. For such nuclei, the beyond mean field methods 
maybe required \cite{BBH.06,LNVMLR.09}, as correlations due to 
configuration mixing and angular-momentum 
projection can affect the relative energies of the various minima.
However, the description of rotational spectra within such models
requires the use of a adjustable scaling factor for the 
moments of inertia, as time-odd mean fields are neglected in the 
current realizations of these methods \cite{LNVMLR.09}. In 
addition, the calculations of odd-mass nuclei may be problematic 
in such approaches since they require potential energy surfaces
over a substantial deformation space (including also triaxiality)
for a fixed blocked one-quasiparticle configuration. However, 
our experience \cite{AO.13,AS.11} tells that in medium and heavy 
mass systems it is frequently impossible to get a convergent 
solution for such configurations even in local minima; the problem 
will become even more numerically unstable when constraint(s) on 
collective coordinate(s) are involved. In addition, the tracing of 
blocked one-quasiparticle states of a given structure over a large 
deformation space is highly non-trivial problem in the methods 
based on the variational principle when triaxiality is involved.

\vspace{-0.2cm}
\begin{ack}
Useful discussions with R.\ V.\ F.\ Janssens and his valuable suggestions
for manuscript are gratefully acknowledged. 
This work has been supported by the U.S. Department of Energy under the 
grant DE-FG02-07ER41459. This research was also supported by an allocation 
of advanced computing resources provided by the National Science Foundation. 
The computations were partially performed on Kraken at the National Institute 
for Computational Sciences (http://www.nics.tennessee.edu/).
\end{ack}

\end{document}